# Gap Risk KVA and Repo Pricing:
## An Economic Capital Approach in the Black-Scholes-Merton Framework

Wujiang Lou[1]

1st draft March 8, 2016; Updated 10/18/2016

**Abstract**

Although not a formal pricing consideration, gap risk or hedging errors are the norm of derivatives businesses. Starting with the gap risk during a margin period of risk of a repurchase agreement (repo), this article extends the Black-Scholes-Merton option pricing framework by introducing a reserve capital approach to the hedging error's irreducible variability. An extended partial differential equation is derived with two new terms for expected gap loss and economic capital charge, leading to the gap risk economic value adjustment and capital valuation adjustment (KVA) respectively. Practical repo pricing formulae is obtained showing that the break-even repo rate decomposes into cost of fund and economic capital charge in KVA. At zero haircut, a one-year term repo on main equities could command a capital charge as large as 50 basis points for a 'BBB' rated borrower.

---

[1] The views and opinions expressed herein are strictly the views and opinions of the author, and do not reflect those of his employer and any of its subsidiaries. The author thanks reviewers and the editors for valuable comments and suggestions. This paper was previously circulated under the title of "Capital pricing during margin periods of risk and repo KVA".

# 1. Introduction

Option pricing lays its foundation on the construction of a trading portfolio in both underlying stock and money market investment such that the option is dynamically hedged. A short option on a stock modeled on geometric Brownian motion, for example, can be self-financed and hedged perfectly so that there is no hedging error. The option and its hedging portfolio therefore contribute zero market risk capital to its trading book. When the stock price is modeled as a jump diffusion process, the jumps in the option price can't be hedged. Consequentially the delta-hedged option portfolio has a hedging error. By assuming non-systemic jumps and the Capital Asset Pricing Model (CAPM)'s applicability, Merton (1976) shows that this error will *average* out to zero and derives a mixed differential-difference Black-Scholes type equation and an option pricing formula with lognormally distributed jumps. Knowing the challenges in CAPM's empirical testing, Merton resorts to the Arbitrage Pricing Theory to derive the same, thanks to the hedging portfolios diversification argument. The hedging error's variability (and thus risk capital) has no pricing impact. In incomplete market theories developed later, a hedging strategy can be optimized to reduce the variance of the hedged portfolio (Duffie and Richardson 1991, Karoui and Quenez 1995). The residual variance, however, had never been a formal part of pricing considerations until the multifaceted post crisis reform of the regulatory risk capital framework arrived.

While a quick and practical response is to incorporate the strengthened regulatory capital requirements into existing derivatives pricing and valuation models, a fundamental revisit of derivatives pricing theories is in order. Post-crisis derivative pricing and risk management research has in fact advanced into a new regime where efforts from both academia and the industry have focused on studying *secondary* risk factors or frictions such as counterparty risk, funding, liquidity, and capital cost. These under-appreciated factors become sources of arbitrage barriers and challenge the foundation of the no-arbitrage belief. While Hull and White (2014) argue that the risk neutral price remains the best estimate of derivative's fair value, Crepey (2011) questions whether the "one-price" rule is still valid and discusses many subtleties and recognize the need of redeveloping models from scratch.

Naturally, a revisionist approach is to expand the classic option hedging setting to accommodate these costs and re-derive the celebrated Black-Scholes-Merton partial differential equations (PDE). Burgard and Kjaer (2011, 2013) incorporate counterparty credit risk into the PDE framework[2] and propose to treat own default risk hedging errors as part of a derivatives desk's funding strategies and capture funding costs under FVA and DVA. The hedging errors are, however, results of an assumed exogenous recovery rate, for example 40%, and an outdated risk-free close-out amount calculation. Lou (2013) shows the hedging error can be eliminated by recognizing the endogenous recovery rate of par when the derivative economy is properly segregated such that, ignoring the gap risk, the residual cash flow of the economy can pay off the debt[3] issued to finance the derivative and its hedges. Assuming the market value recovery, which is more closely aligned to current close-out amount protocol, an economic neutral hedge by the liability-side counterparty is presented that serves as an extension of the Black-Scholes-Merton framework to uncollateralized derivatives (Lou 2015).

While counterparty credit risk and funding costs are met with CVA (credit valuation adjustment) and FVA (funding valuation adjustment), whether cost of capital should be accounted as a cost of business or a new valuation adjustment becomes a heated topic. Green, Kenyon and Dennis (2014) add a regulatory capital term to Burgard and Kjaer's PDE to introduce capital valuation adjustment (KVA), which Prampolini and Morini (2015) find that replacing the junior bond with equity would also arrive at.

In this article, we set out to model and price hedging error associated with gap risk incurred during a margin period of risk (MPR), or MPR gap risk. Margin period gap risk is common and pronounced in securities financing transactions and centrally or non-centrally cleared OTC derivatives. Specifically, we define the MPR gap risk as the collateral market price gap decline during the MPR resulting in loss to repo principal after borrower default settlement. For all practical purposes, we consider the MPR gap risk to be unhedgeable, to have a non-reducible variance, and to be fully warehoused. In the standard market risk capital framework, capital is then required and cost of capital will occur and has to be part of pricing and valuation of derivatives.

---

[2] Counterparty's default risk is fully hedged by shorting the counterparty's zero recovery zero coupon bond, which has no diffusion component in its price dynamics. So counterparty credit risk is in fact not completely hedged should we consider the credit spread risk.

[3] The funding cost on issued notes could be made higher to compensate for the gap risk, including inefficiencies in default settlements.

The main contribution of this article is to introduce a measure of economic capital corresponding to the hedging error and build capital financing cost into the Black-Scholes-Merton pricing framework. By choosing securities financing transactions where sufficient levels of haircuts essentially eliminate counterparty risk and derivatives funding risk, economic capital impact on pre-trade pricing and post-trade valuation can be well isolated and analyzed. Section 2 expands the Black-Scholes option economy for securities financing economy, its funding structure and hedging instruments. Replication portfolio and cash flow analysis are set up with the self-financing condition derived. Section 3 defines economic capital and capital valuation adjustment and links up with the repo economic model (Lou 2016) and derives repo fair value and break-even spread formulae. Section 4 shows repo pricing with examples. Section 5 concludes.

## 2. Modeling a Segregated Securities Financing Economy

Suppose a fictitious bank B (lender) and its customer C (borrower) enter into a repurchase agreement (repo) where the bank lends an amount $N_p$ at interest rate $r_p$ to C. C, in turn, sells stock to B at haircut $h$ and will purchase it back at time $T$. Let $V_t$ be the pre-termination repo fair value. Write $V_t = N_p + v_t$, where $v_t$ is the fair value of the embedded derivative in the repo, likely positive or negative. The bank has in place a hedging strategy $\Delta_t$ in the underlying stock which is traded at price $S_t$. Economically, the reverse repo is a bilateral lending agreement that can be treated as a bilateral defaultable derivative on the stock with counterparty risk. Our intention is to segregate this lending activity into a standalone economy so as to isolate the hedging error.

The economy operates within a capital market consisting of three classes of investors: the equity investor, the capital financer, and the liquidity provider, each having different risk preferences. For example, when underwriting a loan, these investors take on the equity, mezzanine (mezz), and the senior tranches respectively, which are cut to fit the expected loss (EL) of the loan, the unexpected loss (UL) corresponding to a given q-tile, and the remaining size of the loan. A loan for instance can be truncated into a 3% equity, 15% mezz, and 82% senior pieces corresponding to 3% EL and 15% UL on 18% VaR given 99.9 percentile confidence interval.

The liquidity provider looks for opportunities to invest its cash in near credit risk-free products for a finite term earning interest at the rate $r_f$. As the credit risk premium is minimal, the spread over the risk-free rate $r$ reflects the cost of fund (CoF) or liquidity premium of the investment. The equity investor takes any residual loss or gain of the loan in the form of interest margin or excess spread[4]. The capital financer demands for the use of their capital a continuous dividend payout at a yield of $r_k$, while standing by to absorb losses beyond EL.

A 1 year term loan is now tranched in size at C for capital (mezz), SR for liquidity provider, and 1-C-SR for equity, the loan originator's pricing sheet can be simply, *NetCpn = C\*$r_k$ + SR\*$r_f$ + IntMargin,* where C is set to unexpected loss, 1-C-SR equals to expected loss (EL), expressed as percentages of the lending amount. Intuitively the earned interest on the loan needs to cover expected loss, capital financing cost, and the liquidity cost. NetCpn is nominal coupon netted off origination fee and other considerations such as operating costs and profit margins. The IntMargin is supposed to be sufficiently large to cover EL and is held by the equity investor.

Such a market structure, while rather foreign to derivatives pricing literature, is quite standard in insurance industry. In life insurance policy underwriting, for instance, insurers collect sufficient insurance premium to cover the policy pool's projected expected loss in the form of an economic reserve. Initial equity and retained earnings are captured in a special purpose financing vehicle that's used to pay any deficiency of the economic reserve. And a redundant reserve account is partially or fully funded that is last to be tapped into to pay insurance claims, much like the senior piece of the loan syndication.

In the classic Black-Scholes-Merton (BSM) option pricing framework, there is no need for such a capital market setup, as the dynamically replicated option portfolio as a whole has zero expected loss and zero unexpected loss (Shreve, 2004). In fact, we have zero equity tranche since EL=0, and mezz tranche as UL=0, and zero senior tranche (SR=0) as it is self-financed and there is no need for a liquidity provider. One could anticipate that when the secondary risk factors are considered rather than ignored, EL and UL would exist and need to be included in the option pricing framework. The liquidity provider steps in because funding is no longer risk free. As an exploratory note, we adopt the BSM framework for securities financing transactions where funding and capital factors outweigh primary risks such as stock prices.

---

[4] Interest margin is a loan terminology while excess spread is a collateralized loan obligation (CLO) terminology. A loan syndicated with three classes of investors can be considered as a single asset CLO.

**2.1 Accounts Setup**

We start with introducing a number of accounts associated with the bilateral defaultable option economy (Lou 2013) and add other necessary components to the economy.

- **Bank account**: The segregated economy's only investment option is a cash deposit account, with balance $M_t \geq 0$, earning the risk-free rate $r(t)$.
- **Stock financing account**: Stock financing is accessible in either repo or sec lending form. Assuming the reverse repo is long delta, so shares are borrowed from a sec lender to hedge the repo.

Let $h_s$ be the haircut such that the economy needs to post cash collateral of $L_s=(1+h_s)\Delta S$ to the sec lender, $h_s > 0$. The sec lender gives the borrower an interest rebate at a rate of $r_s$ while the borrower pays a manufactured dividend if any to the lender. The difference between the rebate and the risk-free rate is stock's borrowing cost. The margin account is revaluated and margined daily.

- **Repo principal account**: This account represents the funding source of the lending amount of $N_p$. This is from a liquidity provider who charges a cost of fund $r_f$ for the use of liquidity.
- **Economic capital reserve account**: This is a segregated reserve account, with its sole purpose to absorb any potential losses should the borrower default and collateral upon liquidation is not sufficient to pay off the loan from the liquidity provider. The segregated economy requires a reserve balance $N_c \geq 0$ deposited from the capital financier and is responsible for making dividend payment at the rate of $r_k$. The reserve is set aside in a separate bank account, earning interest at $r$.

- **Debt account**: The segregated economy's unsecured cash raising option for the remaining cash flow needs is a short term debt issued and recorded in a debt account with balance $N_t \geq 0$. The debt account has a unit price or par and is rolled at a short rate $r_N(t)$, $r_N(t) > r(t)$.

In Lou (2013) the pair of ($M_t$, $N_t$) is used to capture the funding asymmetry between deposit and borrowing, satisfying $M_t \cdot N_t = 0$, so that cash flow generated in the economy would be used to pay down the debt first since $r_N > r$. Here $M_t \cdot N_t = 0$ is not enforced. The residual cash flow of the transaction is an eligible claim paying resource for this class of notes.

- **Repo margin account:** Repo terms are governed by the Master Repurchase Agreement (MRA) published by the Bond Market Association or the Global Master Repurchase Agreement (GMRA) by the International Capital Market Association (ICMA). From counterparty credit risk mitigation point of view, repo principal $N_p$ is collateralized by the shares of stock.

Derivatives based securities financing transactions such as Total Return Swap (TRS) or Credit Default Swap (CDS) are governed by ISDA Master Agreement where the derivatives mark-to-market is collateralized under the Credit Support Annex (CSA). In securities financing transactions effected by TRS or CDS, the asset bought separately by the lender serves to collateralize its lending.

- **Derivative collateral account:** Under both MRA and GMRA, a repo's accrued interest is part of the daily margin while the present value of the funding interest is not. When it is placed in a trading book, the pv of the repo interest is therefore uncollateralized. Following the liability-side pricing theory (Lou 2015), a fictitious collateral account $L_t$ for $\upsilon_t$. The account $L_t$ will earn the borrower's senior unsecured rate $r_c$ if $\upsilon_t \geq 0$, or the bank's own rate $r_b$ if $\upsilon_t < 0$, i.e., the earned rate on $L_t$ is $r_e(t) = r_b(t)I(\upsilon(t) < 0) + r_c(t)I(\upsilon(t) \geq 0)$.

With derivatives based SFTs, the same collateral account $L_t$ is used here but it pays only the risk-free rate $r$ when a full CSA is in place. Obviously if there is no CSA, the counterparty's rate applies. Let $L_t$ denote the market value of collateral for $\upsilon_t$, posted by party C to B if positive, or by B to C if negative. For simplicity, we assume collateral is full, i.e., $\upsilon_t = L_t$.

Collateral for $V$ is therefore bifurcated into a repo margin account that handles the repo principal and the derivative collateral account. The economy only accepts or posts cash collaterals at zero haircut, and pays or receives $r_e$ or $r$ on the balance $L_t$. Furthermore the collateral process is efficient, prompt and never over-collateralized.

All rates are short rates and determined exogenously. $r_b$ and $r_c$ carry credit risk, while $r_f - r$ is a funding basis similar to the bond-CDS basis. $r_N$ is a placeholder and it could be different from $r_b$ if the bank raises money other than from its senior unsecured rank.

## 2.2 Wealth and Financing Equation

Having listed the economy's various accounts, we now proceed to define the wealth of the SFT economy denoted by $\pi_t$, $t \leq T$.

The repurchase agreement could terminate prior to its scheduled maturity on either party's default. To facilitate discussion of economic capital, we assume repo collateral (the purchased securities) is segregated and not rehypothecated. The borrower can reclaim its securities should the lender default so that the lender does not benefit from its own default. It follows that only the borrower's default needs to be considered. This is consistent with standard loan analysis.

Write the economy's wealth $\pi_t$ as follows,

$$\pi_t = M_t + (1 - \Gamma_t)(V - \Delta S + L_s - L - N - N_p), \qquad (1)$$

where $\Gamma_t = \mathbf{I}(\tau \leq t)$ is C's default indicator, $\tau$ the default time of C. Accounts excluding the cash in the bank account are conditional to no termination, i.e., $\Gamma_t = 0$, so that all relevant quantities are to be understood as pre-default values. To shorten the formula, all $t$-subscripts have been dropped. Following the reduced form modeling approach, we further assume the default time is a stopping time having an adapted intensity process λ defined in the usual probability triplet *(Ω,P,F)*. $M_t$ and $N_t$ are non-negative adapted stochastic processes.

At *t=0*, the wealth reduces to $\pi_0 = M_0 + h_s \Delta_0 S_0 - N_0$. Let $M_0=0$ and $N_0=h_s\Delta_0S_0$ so that $\pi_0 =0$, i.e., the initial capital of the economy starts out at zero. Note that $h_s\Delta S \geq 0$ as when *Δ>0*, $h_s \geq 0$ for sec lending and if *Δ<0*, $h_s \leq 0$ for repo. If the debt account is pegged to the residual stock financing amount, $N_t=h_s\Delta_tS_t$, the wealth of the pre-default economy reduces to the balance of the bank account, $\pi_t=M_t$.

For *τ>t≥0*, during the normal course of business, the bank pursues a trading strategy to hedge the embedded derivative and performs all necessary funding and credit support functions stipulated by the margin agreement. Excess cash is deposited into the bank account, debt, if any, is serviced and rolled as needed. Interests are collected and paid.

Specifically, over a small interval of time *dt*, on the hedge front, shorting or rebalancing dΔ more shares of stock at the price of $S_t+dS_t$ will see cash inflow of $d\Delta_t(S_t+dS_t)$ amount. Since stock shares are borrowed, additional money has to be posted to the sec lender. The debt account pays interest amount $r_N N_t dt$ and rolls into new issuance of $N_t+dN_t$. The bank account accrues interest amount $rM_tdt$. On the collateral side, party C posts additional collateral amount $dL_t$ in cash while being paid of interest amount $r_e L_t dt$. The repo principal account pays out $r_f N_p dt$ and the EC reserve account dividends out an amount of $r_k N_c dt$ while receiving interest of $rN_c dt$. Any increase in repo principal $dN_p$ is funded by the liquidity provider.

The wealth equation is written with all default effects implicitly built into the bank account. If a default happens before *T*, i.e., *τ<T*, trades will have been settled without delay and resulting cash flow will be swiped into the bank account which becomes the only account active till *T*. $M_t$ may exhibit a jump at *τ* as a result of default settlement. Put everything together, the economy's pre-default financing equation, *t<min(T, τ)*, follows,

$$dM = rMdt + (1-\Gamma_t)(r_p N_p dt + d\Delta(S+dS) + dN - r_N Ndt - r_f Npdt \\ - (r_k - r)N_c dt + dL - r_e Ldt - dL_s + r_s L_s dt) \qquad (2)$$

Noting that $L_s=(1+h_s)\Delta S$ and $L=\upsilon$, collect terms to arrive at

$$dM = rMdt + (1-\Gamma_t)((r_p - r_f)N_p dt + d\upsilon - \Delta dS - (r_k - r)N_c dt - r_e \upsilon dt + \bar{r}_s \Delta S dt) \qquad (3)$$

where $\bar{r}_s = r_s(1+h_s) - r_N h_s$ is the economy's effective stock financing rate.

Upon the borrower's default at τ, the hedges can be unwound without loss as stock trades in established exchanges and counterparty risk is negligible. The stock financing account for the delta hedge also unwinds promptly with the overcollateralized amount of $h\Delta_\tau S_\tau$ returned to pay down the issued note $N_\tau$. So no net cash flow arises from stock hedge account $\Delta_\tau S_\tau$, stock financing account $L_s$, and the issued debt account $N$. Furthermore the embedded derivative $\upsilon_t$ offsets with the derivative collateral account $L$.

On the repo, the lending amount at time τ is $N_p(\tau)$, with a collateral market value of $N_p(\tau)/(1-h)$. Here we assume that prior to the default time, i.e., at τ-, there is no missed margin. Effectively we treat the first missed margin call same as default. This is a simplified default timeline for modeling purposes. See Andersen, Pykhtin, and Sokol (2016) for a nice discussion of the margin and default timeline. Assuming that stock collateral is liquidated at the end of the margin period of risk $u$ at a liquidation discount $g$ representing a market liquidity premium, repo margin account would settle at an amount of $\min(N_p(\tau), \frac{1-g}{1-h}\frac{S_{\tau+u}}{S_\tau}N_p(\tau))$. The settlement amount could result in a shortfall to the repo principal amount. The lender's loss is given by

$$l(\tau + u) = (1-R)N_p(\tau)(1 - \frac{1-g}{1-h}\frac{S_{\tau+u}}{S_\tau})^+, \qquad (4)$$

where $R$ is the applicable recovery rate of party C.

Between default time τ and loss settlement at τ+u, the interest to the liquidity provider could keep on accruing. Accrual is ignorable for now as the accrual period is capped by MPR which is short, typically 5 business days for repos. Dividend to the capital financer can be suspended at the point of default, awaiting for default settlement. So the only cash flow during the MPR is the bank account accruing at *r*.

As all accounts match up except the bank account, the wealth of the economy prior to completion of the default settlement is the balance of the bank account. If a loss has occurred, it will be deducted from the balance. If the balance should show a negative value, the segregated reserve account $N_c$ will be drawn to cover the shortfall. To the extent that loss does not exceed $N_c$, the principal amount $N_p$ is without loss and fully returned to the liquidity provider. $N_p$ could still suffer loss, albeit at a negligible tail loss probability, say 0.1% when the confidence interval for capital is set up at 99.9%. Here we assume that this tail loss is priced in $N_p$'s interest rate and absorbed by the liquidity provider.

## 2.3 PDE Derivation

Including the lagged default settlement, the differential wealth equation at time *t*, $t \geq u$, can be written as follows,

$$d\pi_t = r\pi_t dt - d\Gamma(t-u)l(t)$$
$$+ (1-\Gamma(t))((r_p - r_f)N_p dt + d\upsilon - \Delta dS - (r_k - r)N_c dt - r_e \upsilon dt + \bar{r}_s \Delta S dt) \quad (5)$$

The *dΓ* term records default cash flow settlement at time *t* if *t=τ+u*. Assume that the stock price follows a geometric Brownian motion with its real world return *μ* and volatility *σ*, $dS = \mu S dt + \sigma S dW$. Applying Ito's lemma to $\upsilon_t$, setting up delta hedge, and assuming deterministic short rates and default intensity, it follows,

$$d\pi_t - r\pi dt = -d\Gamma(t-u)l(t) + (1-\Gamma_t)\lambda El(t)dt$$
$$+ (1-\Gamma_t)(\frac{\partial \upsilon}{\partial t} + \bar{r}_s S \frac{\partial \upsilon}{\partial S} + \tfrac{1}{2}\sigma^2 S^2 \frac{\partial^2 \upsilon}{\partial S^2} - r_e \upsilon + (r_p - r_f)N_p - (r_k - r)N_c - \lambda El(t))dt \quad (6)$$

where $El(t) = E_t[l(t+u)]$. The second term on the right hand side is a compensator to the first term, assuming that the loss at *t+u* is independent from the default, i.e., no specific wrong way risk. Setting the *dt* term on the right to zero results in the following PDE,

$$\frac{\partial \upsilon}{\partial t} + \bar{r}_s S \frac{\partial \upsilon}{\partial S} + \tfrac{1}{2}\sigma^2 S^2 \frac{\partial^2 \upsilon}{\partial S^2} - r_e \upsilon + (r_p - r_f)N_p - \lambda El(t) - (r_k - r)N_c = 0 \qquad (7)$$

Equation (7) is the governing PDE for the fair price of the repo embedded derivative and thus repo fair value. The first four terms on the left hand side are the same as in the liability-side pricing of derivatives (Lou 2015). The fifth term is the repo financing earned income, the sixth is the charge off or time decay of the expected loss during the MPR, and the last term is the cost of financing economic capital due to the gap risk. The last two terms are absent from typical FVA PDEs, e.g., Burgard and Kjaer (2013), as the gap risk is not considered there. The sixth term distinguishes from other KVA PDEs such as Green et al (2014) where a direct charge of cost of capital is exerted on regulatory capital without modeling gap risk and its economic value.

The net wealth of the segregated economy, $\pi_t$, represents a hedging error resulting from the gap loss *l(t)* upon a jump-to-default event. In the next section, we pinpoint the economic capital corresponding to the hedging error for an SFT economy, which enters the above PDE but has yet to be specified.

**3. Measuring and Managing Hedging Error with Economic Capital**

Now suppose that the embedded derivative fair value process $\upsilon_t$ satisfies the PDE (7), the hedging error term (6) becomes

$$-dA_t = d\pi_t - r\pi dt = -d\Gamma_{t-u} l(\tau + u) + (1 - \Gamma_t) El(t) \lambda dt. \qquad (8)$$

For time s>0, $dA_t$ can be integrated[5] to get,

---

[5] Because of the delayed settlement, the jump term needs to be integrated from *u* to *s+u*, while the survival term from *0* to *s*.

$$A_s = \Gamma_s l(\tau) - \int_0^s (1 - \Gamma_t) El(t) \lambda dt. \tag{9}$$

Assuming default settlement loss is independent of the jump-to-default event, it can be easily verified that $E[A_s]=0$ and $A_s$ is a martingale with $A_0=0$ as, for $s>t>u$, conditional on no default at time $t$, $t<\tau$,

$$A_s - A_t = I(t < \tau \leq s) l(\tau) - \int_t^{s \wedge \tau} El(y) \lambda dy). \tag{10}$$

$A_s$ can be interpreted as the net hedging loss accumulation up to time $s$, excluding its risk-free growth in the bank account. A natural way to look at the economic loss and capital requirement is to consider the terminal loss $A_T$.

### 3.1 Determining Economic Capital

Let's examine $A_T$'s VaR measure given a q-tile,

$$VaR_A = \inf\{x \in [0, +\infty) : \Pr(A_T > x) \leq 1 - q\}. \tag{11}$$

where the tail distribution of $A_T$ is given by

$$\Pr(A_T > x) = \int_0^T \Pr[l(\tau) > (x + \int_0^\tau El(t) \lambda dt)] dP(\tau) \tag{12}$$

Because $E[A_T]$ is zero, economic capital is the same as $VaR_A$, i.e., $N_c=VaR_A$. Obviously, $VaR_A$ can be defined at a forward time t, assuming party C has not defaulted yet by then.

$VaR_A$ is for a fixed time, maturity $T$, while the reserve account $N_c(t)$ is set up starting from time zero. A simple adjustment is to discount it back, i.e., $N_c(t)=VaR_A * \beta(T)/\beta(t)$, where $\beta(t) = \exp(-\int_0^t rds)$. This definition naturally fits with hedging error minimization schemes at some fixed time horizon such as Duffie and Richardson (1991).

Alternatively hedging errors can be examined and minimized locally. Multiply the risk-free deflator $\beta_t$ to equation (8) and integrate from $t$ to $T$ to arrive at, $\beta_T \pi_T - \beta_t \pi_t = \int_t^T \beta_s dA_s$. Because the deflator is continuous and $A_s$ is a martingale, the discounted hedging error is also a martingale. On a no-default path where $\Gamma_s=0$ and $d\Gamma_s=0$ for $t \leq s \leq T$, the path integral gives

$$\beta_T \pi_T - \beta_t \pi_t = \int_t^T \beta_s El(s+u)\lambda ds. \tag{13}$$

This shows that the discounted wealth of the economy accumulates continuously over time on a no-default path, at the rate of default intensity times the expected loss during the MPR.

On a default path with default time $\tau<T$, it jumps due to default settlement that could lead to a loss, and the path integral becomes

$$\beta_T \pi_T - \beta_t \pi_t = \int_t^\tau \beta_s El(s+u)\lambda ds - \beta_\tau l(\tau+u) \tag{14}$$

Note that we always have $\beta_T \pi_T = \beta_\tau \pi_\tau$ for $\tau<T$.

Now we can define the loss of economic value at time t for the full remaining duration of the economy and its VaR as follows,

$$\hat{\pi}_t = \pi_t - \frac{\beta_T}{\beta_t}\pi_T$$

$$VaR_t = \inf\{x \in R : \Pr(\hat{\pi}_t > x) \leq 1-q\}. \tag{15}$$

Because $E[\hat{\pi}_t]$ is zero, $N_c(t)=VaR_t$. This definition is simpler in form and used for equation (7). When a loss is realized and the wealth is not sufficient to pay the loss, money in the reserve account will be drawn to cover loss. On an expectation basis, the wealth growth is sufficient to meet the loss, so economic capital is indeed for unexpected losses.

If the asset price dynamics do not depend on the default of the borrower, the tail probability of $\hat{\pi}_t$ becomes

$$\Pr(\hat{\pi}_t > x) = \int_t^T E[I(\hat{\pi}_t > x) | \tau] dP_t(\tau), \qquad (16)$$

where $dP_t(\tau)$ is probability of default at time $\tau$ as seen on time $t$, i.e., forward default curve,

$$E[I(\hat{\pi}_t > x) | \tau] = \Pr(l \geq b_\tau),$$
$$b_\tau = \frac{\beta_t}{\beta_\tau} x + El \int_t^\tau \lambda \frac{\beta_s}{\beta_\tau} ds \qquad (17)$$

## 3.2 New Valuation Adjustments GAP_EVA and KVA

With economic capital defined, the PDE (eqt.7) is fully specified. Applying Feynman-Kac theorem and noting $\upsilon(T)=0$ arrive at the pricing formulae,

$$\upsilon(t) = E_t \left[ \int_t^T e^{-\int_t^s r_e du} (s_p N_p(s) - s_k N_c(s) - \lambda El(s)) ds \right] \qquad (18)$$

where $s_p = r_p - r_f$ is the liquidity spread and $s_k = r_k - r$ is the capital charge spread. Let $\upsilon^*$ be the risk-free price without consideration of counterparty risk and the gap risk on the MPR,

$$\upsilon^*(t) = E_t^Q \left[ \int_t^T e^{-\int_t^s r du} s_p N_p(s) ds \right]$$

Let $U(t)=\upsilon^*(t)-\upsilon(t)$ be the valuation adjustment to $\upsilon^*$, given by

$$U(t) = CRA + GAP\_EVA + KVA,$$
$$CRA = E_t^Q \left[ \int_t^T (r_e - r) \upsilon^*(s) e^{-\int_t^s r_e du} ds \right],$$
$$GAP\_EVA = E_t^Q \left[ \int_t^T \lambda El(s) e^{-\int_t^s r_e du} ds \right], \qquad (19)$$
$$KVA = E_t^Q \left[ \int_t^T s_k N_c(s) e^{-\int_t^s r_e du} ds \right]$$

In addition to the counterparty risk adjustment (CRA), we now find that capturing gap risk introduces two new fair value adjustments, namely the gap risk value adjustment (GAP-EVA) and economic capital value adjustment (KVA), reflecting the economic value added and the financing cost of the gap-risk capital, respectively. There is no overlap between CRA and KVA as the latter is dedicated to gap risk in MPR which is not present in CRA. Note that GAP-EVA is new compared to other KVA researches (Green et al 2014), where KVA is based on regulatory capital instead of economic capital as is the case here.

If expected loss *El* and economic capital $N_c(t)$ are computed exogenously, for instance, by means of referencing a standardized schedule, the PDE can be solved by finite difference methods. Decomposition of CRA into bilateral, coherent CVA and FVA can be done by solving the PDE with shifts in each parties' synthetic or cash curves (Lou 2015).

Because KVA itself is part of the fair value, the formulae for KVA is recursive in general. An implementation directly calculating KVA however risks to have open IR01 that could attract capital in the same way as FVA when defined imprecisely.

### 3.3 Repo Formulae

PDE (7) specifies precisely repo fair value's dependence on the stock price, in the same fashion as an option. The general understanding and business practice, however, don't normally treat repos as derivatives because of the presence of significant haircuts, daily margin and covenants, especially for non-cusip based repo financing facilities that would forestall asset price dependency. In fact, if $v_t$ is not a function of *S*, PDE (7) reduces to an ordinary differential equation,

$$\frac{dv}{dt} - r_e v + s_p N_p - s_k N_c - \lambda El(t) = 0 \qquad (20)$$

When $N_p$ and *El* are constant in time, and if $N_c$ is set to a constant, e.g., the average capital during a carrying period, with constant $\lambda$, the repo fair value formula is found by solving equation (20),

$$\upsilon = (s_p N_p - s_k N_c - \lambda El) apv \qquad (21)$$

where *apv* is the borrower's risky discounted annuity. For pre-trade pricing, by setting $\upsilon_t$ to zero, we obtain the break-even repo formula,

$$r_p = r_f + s_k \frac{N_c}{N_p} + \lambda \frac{El}{N_p}. \qquad (22)$$

We now see in equations (21&22) the economic capital charge intuitively. Equation (22) in particular formalizes what street traders have been doing when adding a regulatory capital charge to cost of fund $r_f$, together with an interest margin for profit, to arrive their repo rates. *El* and $N_c$ are provided by a haircut model that considers the joint dynamics of collateral asset and borrower credit, for instance, Lou (2016), where correlation between the asset return and the credit spread is shown to have only marginal impact. This is due to daily margining and a short MPR. Consequently there is not a strong practical need to introduce stochastic default intensity. If the borrower's credit spread is deterministic or stochastic but independent of the asset price process, then EL and EC can be computed as a deterministic function of time *t*, followed by repo fair value and valuation adjustments, by solving ODE (20).

Of course for constant position securities financing transactions, such as a TRS on a fixed number of shares of stock or fixed notional of bond, the full PDE applies.

Alternatively, a repo can be seen as a form of counterparty credit derivatives and be priced following the conventional credit derivatives risk-neutral pricing approach. From a credit derivatives perspective, the lender sells a funded protection to the borrower[6]. Briefly, let *β(t)* be the applicable discount factor, the net present value (npv) of the loss, or default present value (dpv), is given by,

---

[6] If the borrower defaults and the lender suffers a loss after a repo settlement, the borrower's estate can be viewed as if having taken the opposite, i.e., benefitted from not having to pay back the full principal borrowed.

$$dpv = E[\int_0^T \beta(t)dl(t)] = \beta(T)E[l(T)] - \int_0^T E[l(t)]d\beta(t), \qquad (23)$$

The reverse repo is effectively a floating rate note on a rate index, commonly LIBOR for tenors of 1 month or longer. The index part of the repo interest rate gets back to par when discounted at the same index curve. The present value of a unit spread, or annuity denoted by *apv*, is given by $apv = E[\int_0^T \beta(t)Q(t)dt]$, where $Q(t)$ is party C's survival probability and the annuity is computed on a unit notional, suitable for a repo funding facility. The net present value of the repo (npv) is then *npv=1-dpv+s_p\*apv* where $s_p$ is the repo spread.

The pair of dpv and apv are essentially what the conventional credit derivatives pricing technique brings about. For a repo, however, these are not materially relevant, for dpv is very small due to the presence of significant haircuts and apv is very close to *T* when *T≤1* year, unless with extreme short term discount rate and low quality credit counterparty. What matters to repos are cost of fund and how much capital is used at what cost, both of which are missing from the above. This shows again the inability of the risk-neutral pricing theory to capture gap risk and economic capital.

Lou (2016) presents a repo haircuts model and a companion economic capital model linking for the first time haircuts and economic capital to collateral security's price dynamics and borrower's credit spread dynamics. The exposure of an SFT is made equivalent to a secured term loan to a wholesale counterparty. The economic capital model takes into account asset risk, credit risk, wrong way risk, and market liquidity risk. General wrong way risk (WWR) is modeled by correlation between the asset return and the credit spread. The specific WWR is captured by a jump-on-default liquidation discount. Economic capital is defined as the unexpected loss taken either as the VaR or expected shortfall minus the expected loss. Historically estimated double exponential jump diffusion model shows that the regulatory capital undercuts economic capital in a risk-on range of haircuts and overstates economic capital in a low economic risk range.

As expected, the general WWR is weak and negligible, because of the frequent margining and the MPR is short. Consequently there is not a strong practical need to introduce stochastic default intensity. If the borrower credit spread is deterministic, then EL and EC can be computed as a deterministic function of time *t*. Repo fair value and valuation adjustment can be computed straightforwardly. If a dynamic spread such as the log-Ornstein-Uhlenbeck default intensity model is used and is independent of the asset price process, EL and EC can be computed separately and plugged into the fair value formula to arrive at repo fair value.

Note that the PDE is derived assuming a geometric Brownian motion. When stock price follows a jump-diffusion process, the PDE would involve fair value jumps corresponding to jumps in stock prices, and become a partial integral differential equation. The fair value expectation formula and valuation adjustments however remain valid as the jumps are reflected in the probability measure under which the expectation is taken.

For collateralized OTC derivatives, margin period of risk exists and initial margin is called upon as a reserve. PDE (7) still applies, with repo principal term dropped out, once $El$ and $N_c$ for the gap risk are established for the OTC derivatives book. Needless to say, economic and regulatory capital requirements, and thus KVA, must be computed at a book level. The capital pricing analysis could be extended appropriately, though computational concerns naturally follow. These are left for future research.

**4. Repo Pricing and Valuation Examples**

The model presented can be used for pre-trade repo pricing and post-trade valuation. With the former, traders often find themselves negotiating with clients either haircut levels, or repo rates, or both, and sometimes have to go with the flow in the market to be competitive. Naturally, low haircuts would need to be compensated with higher repo rates. Our model establishes exactly such a quantitative link between haircuts and repo rates via economic capital and its capital charge. With the latter, while repos are typically treated on accrual accounting as secured loans for accounting purposes, repo fair valuation with sufficient sophistication has become a pressing need, especially for broker-dealers acting as financing intermediaries between hedge funds and money market funds and treating repos as trading book positions that require fair value accounting. BASEL's new minimum market risk capital requirements specifically ask for measuring repos' counterparty credit risk. In this section, we demonstrate the model's applications.

The sample repo is a one-year term repo earning 60 bp spread (repo rate minus cost of fund) with US main equities as collateral. SPX500 index is used as a proxy for the collateral asset. A double-exponential jump-diffusion model is estimated from daily price history of SPX500 from January 2008 to January 2013 for haircut and economic capital calculation (Lou 2016). It has an annualized diffusion volatility of 24%, up jump probability of 46%, averages 3.2 jumps during an MPR of 10 trading days, average up jump size of 0.59%, and average down jump size of 0.78%. The borrower is a 'BBB' party with its 5 year CDS traded at 250 bp level.

Suppose client inquiries are at three haircuts, 0%, 5% and 10%. repo fair value and valuation adjustments are shown in Table 1. KVA strikes out and dominates other adjustments. As expected, GAP-EVA is not a factor and can be dropped from the repo pricing formula (eqt. 20). CRA (CVA plus FVA) is very small as the counterparty credit risk is already mitigated by repo margining.

Table 1. Fair value and valuation adjustments in running spread (bp) of the sample one year reverse repo on SPX500 under 0, 5%, and 10% haircuts. *r=0.7%, basis 0.5%, 1y CDS 1.88%.*

| HC  | npv*  | CRA  | GAP-EVA | KVA   | npv   |
|-----|-------|------|---------|-------|-------|
| 0   | 59.78 | 0.73 | 0.07    | 55.51 | 3.47  |
| 5%  | 59.78 | 0.73 | 0.01    | 26.98 | 32.05 |
| 10% | 59.78 | 0.73 | 0.00    | 4.09  | 54.96 |

In pre-trade pricing, the break-even earned spread can be solved by setting npv to zero, or equivalently by adding CRA, GAP-EVA and KVA. At 10% haircut, it will be 4.82 bp. If the client wants to trade at a lower haircut, say 5%, the break-even increases to 27.72 bp, exactly showing the tandem effect between haricut and repo pricing.

Intuitively, because repo tenors are short and risks embedded are similar to gap risk in nature, repo valuation given sufficient haircut levels shall reduce to taking the present value of the repo spread. The case of 10% haircut in Table 1 is a good example as its npv of 54.96 bp is close to the spread of 60 bp.

As been made clear, this paper associates KVA to economic capital rather than regulatory capital (RC). Under BASEL III, a credit risk exposure is calculated to the extent the trade haircut is less than the required regulatory haircut. A capital charge could be calculated by applying the firm's return on equity (ROE) target to the regulatory capital, although use of regulatory capital in trade pricing is definitely not BASEL's original intention. Figure.1 shows a comparison of repo mark-to-market with a capital charge collected on EC and RC, when traded at haircuts ranging from zero to 20%. The most noticeable difference lies in the low haircut (thus more risky) region, where MTM with RC is significantly over-valued than MTM with EC, because RC is significantly lower than EC. In particular, at zero haircut, pv with EC is about 3.47 bp while pv with RC is 36.42 bp. This observation is important for derivatives (e.g. TRS) based securities financing which is often conducted at zero haircut. This shows that KVA based on RC could distort true trade economics and valuation.

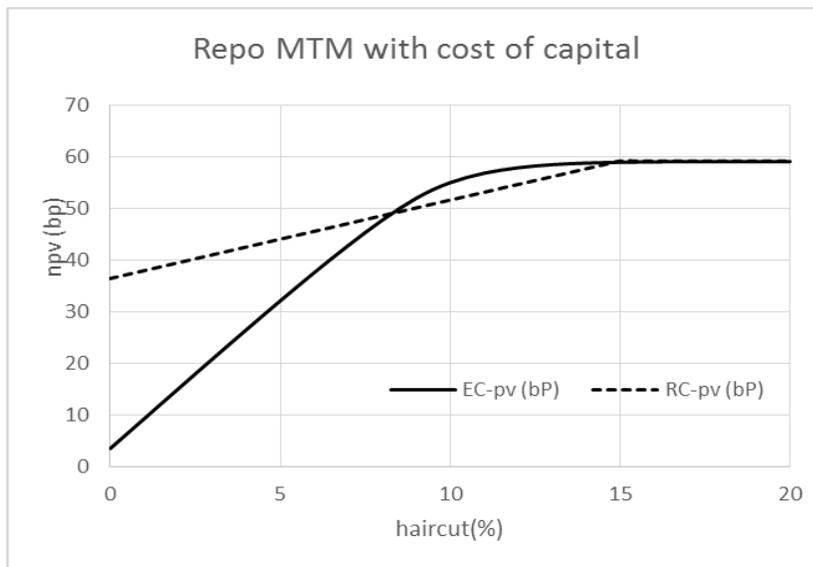

Figure 1. Repo MTM with cost of economic capital (labelled 'EC-pv') compared with regulatory capital (labelled 'RC-pv') vs haircuts. 1y repo, 60 bp repo spread, 10% ROE, and $r_c= 3.1\%$.

Economic capital tends to decay through time as the remaining maturity of the trade shortens[7]. Figure 2 shows projected economic capital (measured with expected shortfall) for the sample repo trade. With zero haircut, EC remains elevated across the full duration, only tapering off as it approaches maturity. At 5% haircut, EC reduces by about half. At 10% haircut, EC starts at 0.72% and declines almost linearly to zero at maturity.

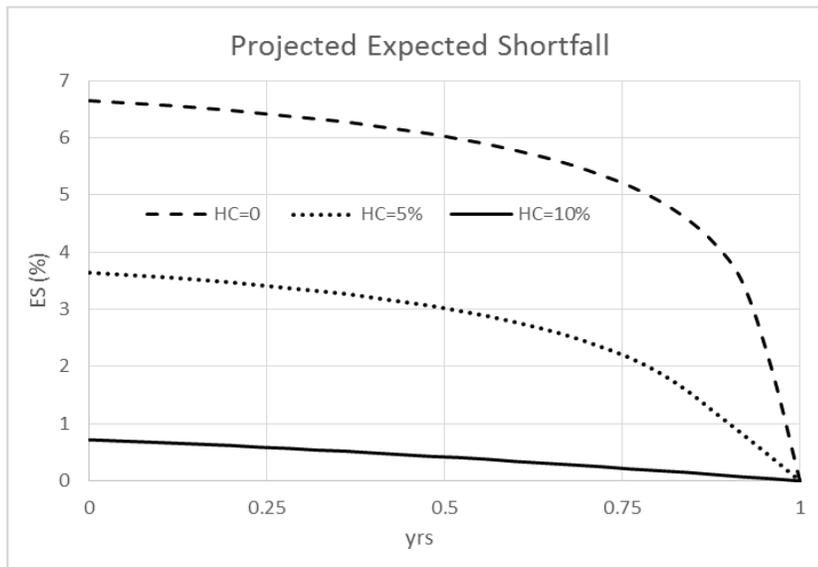

Figure 2. Expected shortfall EC profile for a one year repo on SPX500 main equity under haircuts of 0%, 5%, and 10%. MPR=10 days.

---

[7] Regulatory credit risk capital rule adopts a 1 year floor.

Admittedly the pricing example given above is a standalone repo on stocks[8]. The discussion however is pertinent, for commercial banks and insurance companies typically have books sit on the lending side and don't enjoy any offsetting effect. For a dealer's active repo trading book, netting would kick in and affect EC calculation in principle, but the real effect could be limited as dealers tend to trade repos at high haircuts and run a back-to-back book.

Whether to include cost of capital in accounting fair value is debatable, as it is usually considered a cost outside of the revenue and income stream to which the fair value belongs. Others argue that, since it is charged by a party stepping in, it should be reflected in the fair value per IFRS 13's exit price based fair value measurement. In our view, the more pressing question is what it accounts for economically. KVA should be based on economic capital and account for hedging errors or unhedgeable risks, not *unhedged* risks. A measure of forward economic capital cost can always be calculated and kept as a valuation reserve to be released over time (similarly suggested in Albanese et al 2016), if not formally accepted into accounting fair value.

The computation above shows an ideal case where a second order risk effect outweighs first order. Traditional risk neutral pricing is correct only to the extent that complete hedges can be constructed or hedging errors can be perfectly diversified to nil. Otherwise, economic capital will result and cost of capital has to be evaluated and treated properly. Charging clients for capital cost is not only necessary but also an extension of existing derivatives pricing theories where prices are computed on expectation under a risk neutral measure without consideration of risk capital.

## 5. Conclusion

Recognizing the hedging error during margin periods of risk is unhedgeable, we set out to define economic capital and incorporate the capital financing cost into derivatives pricing and valuation. The Black-Scholes-Merton pricing framework is extended to model the MPR gap risk and capital charge inherent in securities financing transactions. At zero haircut, a one-year repo with 10 day MPR on main equity could command capital charges as large as 50 bp for a 'BBB' rated borrower. Increased haircut reduces capital charges, e.g., to 4 bp at 10% haircut. This shows that for derivatives based funding transactions such as TRS transacted at zero haircuts and OTC derivatives collateral with insufficient haircuts, the cost of economic capital has to be captured in pricing.

---

[8] When the underlying security is a corp bond, the same analytics applies but bond maturity and volatility would have to be considered. Typically, a proxy bond price index of the same maturity bracket (say 3 to 5 years) is picked and EC is computed. See Lou (2016) for more details.

To be more truthful and sensitive to actual economic risks associated with hedging errors or outright unhedgeable risks, we introduce KVA as cost of economic capital (instead of regulatory capital), and discover a new gap-risk added economic value adjustment (GAP-EVA). We also obtain practical repo pricing formulae where the break-even repo rate decomposes into cost of fund, counterparty credit risk/GAP-EVA, and economic capital charge/KVA. With reasonable haircuts, GAP-EVA is very small and KVA dominates other valuation adjustments.

Our approach in establishing economic capital and its financing cost to manage hedging errors can be extended to margined OTC derivatives' MPR gap risk and finds applications in derivatives hedging and valuation in incomplete markets. These are left for future research.